# Deep Neural Network based Wide-Area Event Classification in Power Systems

Iman Niazazari
*Department of Electrical and Biomedical Engineering*
*University of Nevada, Reno*
Reno, USA
niazazari@nevada.unr.edu

Amir Ghasemkhani
*Department of Computer Science and Engineering*
*California State University San Bernardino*
San Bernardino, USA
Amir.Ghasemkhani@csusb.edu

Yunchuan Liu
*Department of Computer Science and Engineering*
*University of Nevada, Reno*
Reno, USA
ycliu@nevada.unr.edu

Shuchismita Biswas
*Department of Electrical and Computer Engineering*
*Virginia Polytechnic Institute and State University*
Blacksburg, USA
suchi@vt.edu

Hanif Livani
*Department of Electrical and Biomedical Engineering*
*University of Nevada, Reno*
Reno, USA
hlivani@unr.edu

Lei Yang
*Department of Computer Science and Engineering*
*University of Nevada, Reno*
Reno, USA
leiy@unr.edu

Virgilio Centeno
*Department of Electrical and Computer Engineering*
*Virginia Polytechnic Institute and State University*
Blacksburg, USA
virgilio@vt.edu

*Abstract*—This paper presents a wide-area event classification in transmission power grids. The deep neural network (DNN) based classifier is developed based on the availability of data from time-synchronized phasor measurement units (PMUs). The proposed DNN is trained using Bayesian optimization to search for the best hyperparameters. The effectiveness of the proposed event classification is validated through the real-world dataset of the U.S. transmission grids. This dataset includes line outage, transformer outage, frequency event, and oscillation events. The validation process also includes different PMU outputs, such as voltage magnitude, angle, current magnitude, frequency, and rate of change of frequency (ROCOF). The simulation results show that ROCOF as input feature gives the best classification performance. In addition, it is shown that the classifier trained with higher sampling rate PMUs and a larger dataset has higher accuracy.

This material is based upon work supported by the Department of Energy National Energy Technology Laboratory under Award Number DE-OE0000911. This work was prepared as an account of work sponsored by an agency of the United States Government. Neither the United States Government nor any agency thereof, nor any of their employees, makes any warranty, express or implied, or assumes any legal liability or responsibility for the accuracy, completeness, or usefulness of any information, apparatus, product, or process disclosed, or represents that its use would not infringe privately owned rights. Reference herein to any specific commercial product, process, or service by trade name, trademark, manufacturer, or otherwise does not necessarily constitute or imply its endorsement, recommendation, or favoring by the United States Government or any agency thereof. The views and opinions of authors expressed herein do not necessarily state or reflect those of the United States Government or any agency thereof.

## I. INTRODUCTION

### A. Motivation

Disruptive events such as line outage, transformer outage, and frequency events occur in power grids from time to time and can lead to temporary or sustained failures with potential expensive replacement or repair cost over time. For instance, according to the report by the U.S. Department of Energy (DOE), transmission transformers failures can lead to the replacement cost between $2 million to $7.5 million [1].

The ice storm of 1998 caused substantial damage to trees with major fault and electrical infrastructure failure events over parts of northern New England, northern New York, and southeastern Canada. Millions were left without power for several days to weeks, leading to nearly 40 fatalities, and the storm damage cost of over $3 billion [2]. The most recent ice storm incident of 2013 affected much of the United States and parts of Canada, resulting in millions of customers without power, 27 deaths and over $200 million in estimated total damages [3]. One of the most prominent examples of failures in delivering a reliable electricity service is the 2003 blackout in the Northeast United States, which was initiated and continued with the occurrence of several faults and assets' failures. At the time, it was the second most widespread outage in history which resulted in power outages to almost 50 million Americans in eight states with an estimated cost of 4 to 10 billion dollars. When the results of the joint U.S.-Canada Power Systems Outage Task Force report was published, four primary reasons for the blackout was pointed out. Among those, one was inadequate situational awareness in the grid [4].

As the cost of failure in energy delivery is far more than the cost of replacing a device, it is of paramount importance to find the cause of problems and fix them before they turn into a catastrophe. This calls for advanced and robust event classification as an essential tool that can provide important information of the grid and the asset conditions, increase the situational awareness of the grid, and enhance the health monitoring of essential and critical assets. Advanced event analytical platform will also benefit electric utilities in many aspects, including increasing situational awareness, increasing assets health monitoring, enhancing power quality report with better root cause analysis of events, scheduling preventive maintenance of assets, decreasing equipment replacement cost, avoiding unexpected outages, decreasing number of maintenance crew utilization, increasing reliability of the system, and increasing the life expectancy of critical assets.

Thanks to the proliferation of advanced metering devices with high sampling rates in power grids, such as phasor measurement units (PMUs), large amounts of measurements provide unprecedented potentials for wide-area monitoring, control, protection, and device-level or systems-level diagnostics applications, such as situational awareness of the grid or health monitoring of grid assets. Thus motivated, this paper aims to develop a robust data-driven event classification platform for classification of events.

*B. Related Works*

Event classification has been studied in previous works [5]-[13]. In [14], six types of transients (including breaker switching, capacitor switching, short circuit fault, primary arc, lightning disturbance and lightning strike fault) are simulated and classified using the wavelet transform. In [15], a spatiotemporal feature representation is leveraged for classifying five electromagnetic transient events (including line switching, capacitor bank switching, fault, lightning, and high impedance faults). The events are simulated in both EMTP and RSCAD and the results are compared in different scenarios such as the effect of different numbers and locations of measurement devices. In [16], the authors used voltage measurements to classify voltage dip events (including fault-induced, transformer saturation, induction motor starting along with non-fault and fault-induced interruptions). In [17], a technique based on the wavelet transform and hybrid principal component analysis is proposed to classify and localize the switched capacitor bank events.

However, existing studies fail in addrssing either two things: they did not use an efficient way to train the model, or they did not study the wide-area real-world events. To tackle these issues, this paper leverages Bayesian optimization to search for the best hyperparameters when training deep neural network based event classifier. In addition, the classification model is trained based on the real events extracted from PMUs streams in the U.S. transmission grids.

*C. Summary of Main Contributions*

The goal of this paper is to develop a wide-area data-driven framework for event classification in transmission power grids based on PMU measurements. Fast and accurate classification of events will lead to a more accurate root cause analysis of failures in the grid, a quicker system restoration after disturbances, and blackouts.

The dataset for this study is from the real-world PMUs of the U.S. transmission grids. The dataset includes four major types of events, namely line outage, transformer outage, frequency event, and oscillation event. Compared to the simulated datasets used in most of the studies, a big challenge in dealing with real-world datasets is due to the missing data. Missing data could be mainly the result of losing the GPS signal or missing communication. To train a good neural network, the missing data needs to be addressed before studying the feature extraction and hyperparameter tuning. A data preprocessing scheme is proposed in this paper to address the issue of missing data.

To perform the data-driven event classification, the use of deep neural network is presented in this paper. Neural network-based methods have several advantages, such as better use of historical trends and patterns in the collected measurements and datasets, less need for human intervention and feature extraction, continuous improvement by adding newly available datasets, and more effectively handling of multi-dimensional and multi-varsity datasets. In this paper, a deep neural network (DNN) is used to fit a model to the training dataset.

A major issue in the training of a DNN on the dataset is how to tune hyperparameters. There are several ways to train neural networks. One way is to manually tune the hyperparameters with trial and error to select the best combination of hyperparameters for achieving the best accuracy. Another way is to apply a random search to find the best model, where the hyperparameters are randomly selected in each iteration until the desirable model performance is achieved. Grid search is another type of hyperparameters tuning which is very similar to random search; however, instead of setting the hyperparameters random values, the hyperparameters are taken the values from a finite number of values. In this paper, Bayesian optimization based hyperparameter tuning is used in the training of the neural network to find the best hyperparameters.

In summary, this paper presents a wide-area event classification in power systems using a deep neural network classifier trained using Bayesian optimization to search for the best hyperparameters. The effectiveness of the proposed method and the subsequent event classification is validated using real-world PMU dataset of the U.S. transmission grids provided by Pacific Northwest National Lab (PNNL) in an anonymous format [18]. The rest of this paper is organized as follows. Section II proposes the methodology of event classification. Section III presents the case studies of this paper. Finally, the simulation results and conclusions are provided in Section IV and Section V, respectively.

II. METHODOLOGY

In this paper, an event classification task is performed using deep neural networks trained with Bayesian optimization for tuning the hyperparameters.

*A. Deep Neural Networks*

Deep learning is a family of machine learning that imitates the way humans learn certain types of knowledge. Deep learning is an important element of data science, which includes statistics and predictive modeling. It is very beneficial in analyzing and interpreting large amounts of data and makes this process simpler and quicker. It has been used in several fields of power system including load forecasting [19], voltage sag estimation [20], wind speed forecasting [21], and cyber-physical power system studies [22].

A deep neural network (DNN), shown in Fig. 1, is a family of artificial neural networks that has multiple hidden layers instead of one single layer. The beauty of DNN is that it finds a mathematical relation between input and output by finding the corresponding networks' weights through several rounds of training even when it is very complicated for humans. DNN incorporates different linear or non-linear activation functions in the layers to help the network to find the corresponding output by calculating the probability of each output [23].

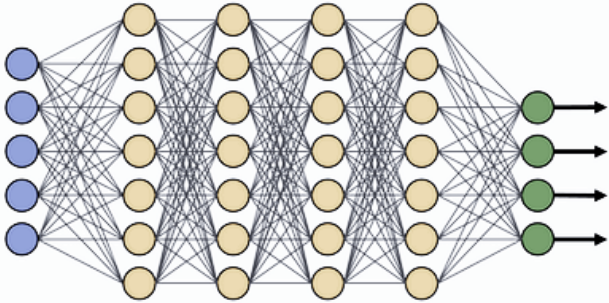

Fig. 1. Structure of a deep neural network

### B. Neural Network Training

The performance of a neural network model is dependant on its hyperparameters values. Therefore, selecting the best combination of the hyperparameters to achieve a well-performed model is essential. The problem with hyperparameter optimization is that it is extremely costly to assess the performance of a set of parameters. This is because we first have to build the corresponding neural network, then we have to train it, and finally, we have to measure its performance on a test set.

Training a neural network can be done in several ways. One way is to manually tune the hyperparameters until with trial and error to select the best combination of hyperparameters for achieving the best accuracy is achieved. One other way is to apply a random search to find the most optimal model. In this method, the hyperparameters are randomly selected in each iteration until the desirable model performance is achieved. Grid search is another type of hyperparameters tuning which is very similar to random search, however, instead of giving the hypermeters random values, they are evaluated based on a finite number of values. Finally, the neural network can be tuned using Bayesian optimization based hyperparameter tuning. In this paper, Bayesian optimization is used in training of the neural network to find the most optimal set of hyperparameters. The flowchart of the algorithm is shown in Fig. 2.

Bayesian optimization aims to construct a surrogate model of the search space for hyperparameters. Gaussian Process is one type of these models. Gaussian Process samples model to maximize the acquisition function (i.e., expected improvement in this paper) to estimate the objective function and evolves during the process to make better predictions. Bayesian optimizer gives us a new suggestion for hyperparameters in a region of the search space that have not been explored before. This process is iterated numerous times until the best hyper-parameters for achieving the best model performance is reached [24].

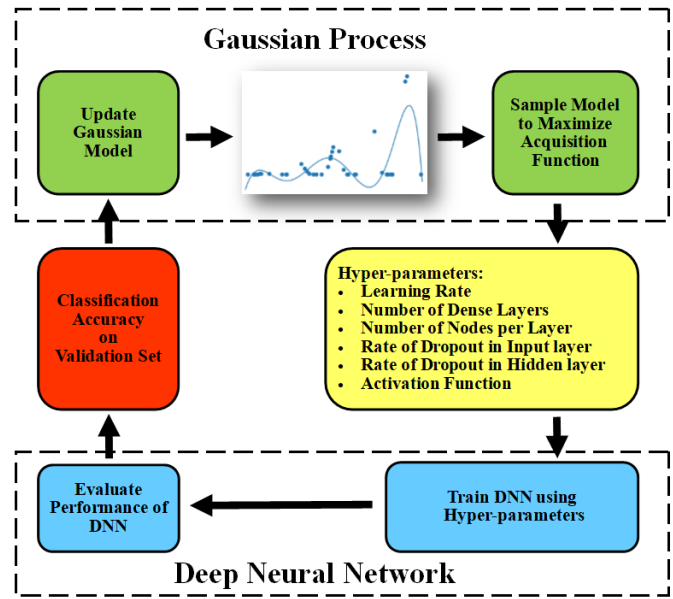

Fig. 2. Flow chart of hyperparameter tuning using the Bayesian optimization

## III. DATASET DESCRIPTION

### A. Dataset

In this section, the case studies for wide-area event classification using the PMU data stream are presented. The dataset for this study is a real-world dataset from the U.S. transmission grids provided by the Pacific Northwest National Laboratory (PNNL) in an anonymous format. The data set are collected from all three interconnections in the U.S. and they are named as A, B, and C by PNNL without revealing the true name of the interconnections. In addition, the network topology, PMU IDs, and the locations are all anonymized by PNNL.

In this paper, we have used the dataset from interconnection B. The dataset is from two years of 2016 and 2017. We have used 12 weeks of data, six weeks of data, from Jan 1, 2016, to Feb 11, 2016, and 6-weeks of data, from Feb 26, 2016, to Apr 7, 2016, for our study in this paper. The events are classified based on the source of the events specified in the event log provided by the PNNL. After processing the events, 127 Line Outage events, 68 Transformer (XFMR) Outage events, 31 Frequency events, and finally 19 Oscillation events, comprising 245 events in total, are selected for our classification training and evaluation. We call these classes, class 0, class 1, class 2, and class 3, respectively. The events have been extracted from the stream of the dataset based on the coordinated universal time (UTC) specified in the event log.

Based on our extensive experiences, we realized the time of event occurrence reported in the event log is not necessarily the exact true UTC time that an event has occurred based on our event detection algorithm which will be presented in future publications. Therefore, we cropped the data sets in an interval of one minute before and three minutes after the reported UTC in the provided event log to make sure that we do not lose any important information in PMU signals during the event extraction process.

From our preliminary studies, we found that it is very challenging to correctly classify each type of events by handpicking the signal features or signatures, such as peak or duration of changes in the stream of voltage or derivative of voltage, current or frequency. Therefore, we propose the use of DNN for automatic signature selection and eventual classification.

The data from Interconnection B includes measurements from 43 unique PMUs with two kinds of sampling rates: 30 frames per second (fps) and 60 fps. One interesting observation is that the 60 fps PMUs report both positive sequence and phase measurements for voltage phasors, but only positive sequence measurements for the current phasor. In the 30 fps PMUs, both phase and positive sequence measurements are present for voltage and current phasors. Another observation is that PMUs with a 60 fps rate gives a better and smother signal compared to PMUs with 30 fps rate. Therefore, we use 23 PMUs with 60 fps for event classification at this stage; however, we compared the classification results with 30 fps PMUs.

*B. Data Preprocessing*

As there are missing data in the PMU dataset, we first try to find the missing ones using other data points in the stream. However, there are few PMUs that the streamed data is completely missing or maybe not reported. For such scenarios, we have not considered them during the training process of DNN. It must be noted that information corresponding to all PMUs is not present for every day in the year. This could be due to multiple reasons: 1) some PMUs may have been commissioned/decommissioned in the course of the year or 2) some PMUs might have been out of service for some days in the year due to planned/unplanned data on particular days. The information obtained is concisely presented via a binary matrix in Fig. 3. Each element in the matrix can take binary values 0 (*yellow*) or 1 (*purple*). Here, the columns represent days in a year, while the rows correspond to PMU ids. Hence element $(i, j)$ in the matrix denotes if data corresponding to the $i$-th PMU is available for the $j$-th day. Of course, this does not specify if a PMU has missing observations on a particular day.

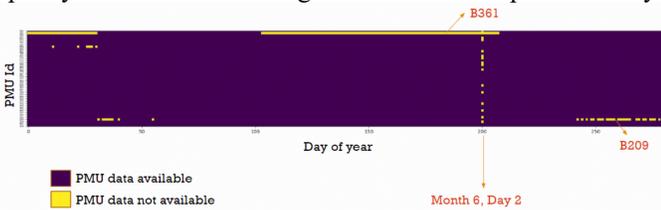

Fig. 3. PMU data availability in 2016 in interconnection B

From our preliminary study, we have realized that it is very difficult to train the DNN solely based on the raw streamed data from PMUs. Therefore, we have decided to calculate the secondary features by taking the gradient of the reported samples and add them to the dataset. Additionally, the dataset is saved for each PMU according to the event name and anonymized PMU ID from the original dataset. Here is the summary of the preprocessing steps:

1. Convert "string" fields in the PMU reported files to "float".
2. Fill out the missing data using a piece-wise linear method for cases where the rate of missing samples to the available samples are low.
3. Calculate the gradient of streamed signals
4. Add new features to the data frame.
5. Save all features for all PMUs.

IV. RESULTS AND DISCUSSIONS

Once the original dataset is preprocessed, we move to the classification step. In this step, using the preprocessed events in the previous section, we train a DNN model to predict the new events. We first train the DNN model by considering all the selected signal (e.g., gradient of voltage magnitude) from all PMUs with the same sampling rate (e.g., 60 fps) over the specified time interval of a 4-minute window of samples (one minute before and three minutes after the event occurrence which is 14400 samples). 75% of events are randomly used for the training set and the remaining 25% of unused events are selected for validation. After training and validating the DNN with 2-dimensional (2-D) input (i.e., PMU signal from different locations over time), we realize that the DNN model does not perform well and the classification metrics, such as the accuracy, are low. The reason is that the performance of DNN models is highly dependent on the size of the training dataset and our 12-weeks data set is not large enough to train a good DNN with 2-D input. Therefore, we have switched to use 1-D input (i.e., PMU signal from each location over time) to the DNN and to consider each PMU separately. Fig. 4 shows the how we separated PMUs to make create 1-D input signal to the DNN.

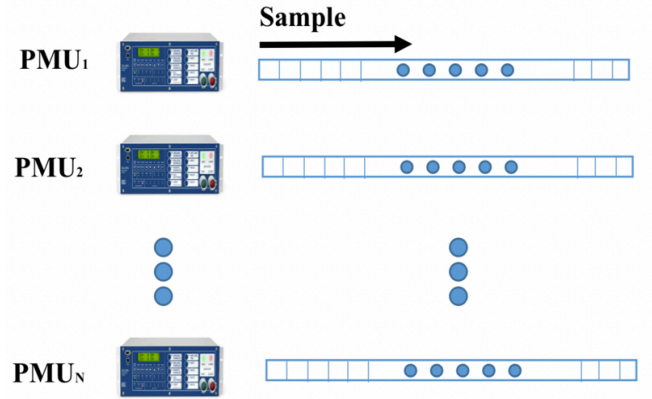

Fig. 4. Representation of events with separated PMU

For running the simulations, the Bayesian optimization is run with 100 calls, each with 30 epochs and batch size of 16. Six hyperparameters are tuned during the training process with the following intervals/choices. Learning-rate, from $10^{-1}$ to $10^{-5}$, the number of hidden layers, from 1 to 8, the number of nodes in each layer, from 20 to 500, rate of dropout in input layer, from 0.4 to 0.9, rate of dropout in hidden layer, from 0.2 to 0.7, and activation function, either 'sigmoid' or 'ReLU'.

*A. Effect of Input Type and Size of Training Dataset*

In this section, the effect of input signal type and size on the model performance is studied. We have studied five different types of inputs to the classifier. We have trained the model on the input with the gradient of positive sequence voltage (GV), the gradient of positive sequence current magnitude (GI), the gradient of positive sequence voltage angle (GV_angle), the gradient of frequency (GF), and rate of change of frequency (ROCOF) features. In addition, to observe the effect of size of training dataset on the event classification, the process is performed for two cases, 6 weeks from Jan 1, 2016, to Feb 11, 2016, and 12-weeks from Jan 1, 2016, to Feb 11, 2016, and from Feb 26, 2016, to Apr 7, 2016. The results are shown in Fig. 5.

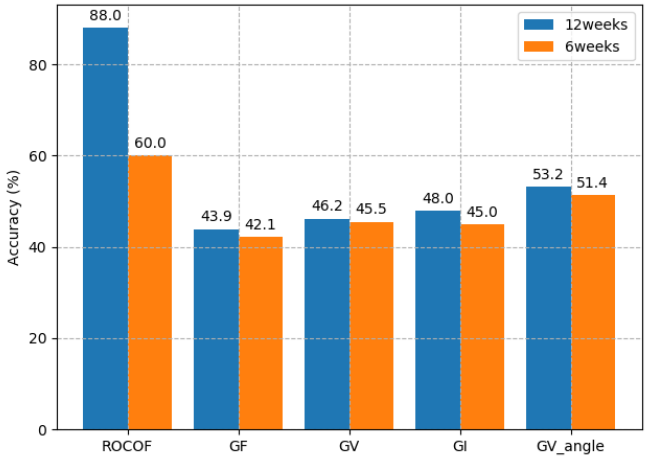

Fig. 5. Effect of input type and size on the event classification

As illustrated in Fig. 5, the best result is achieved for the ROCOF input feature with tuning hyperparameters of learning-rate, number of hidden layers, number of nodes in each layer, rate of dropout in input layer, rate of dropout in hidden layer, and activation function to $1.25 \times 10^{-3}$, 7, 499, 0.42, 0.21, and ReLU, respectively. However, for other features, the accuracy drops significantly. While for Another observation is that as the size of the training dataset increases from 6-weeks to 12-weeks the accuracy improves. This is because the DNN model is trained with more events and more information is provided to the model in the training process. The model accuracy trained on 12-weeks of data with ROCOF feature as input is 88%. This is while for other features the accuracy never passes 55%.

*B. PMU Sampling Rate*

In this section, the effect of the sampling rate of PMUs on event classification is studied. The sampling rate can reveal useful information for grid operators in terms of how capable their measurement equipment is for event classification. Therefore, performing a sensitivity analysis with respect to sampling rate of PMUs and observing the impact on event classification help the utilities and system operators to find out the capabilities of their wide-area measurmeent devices for new tools and applications.

As the training dataset is unbalanced with respect to the distribution of events and classes, by looking at only the accuracy may not give us the best insight into the classifier performance. Therefore, to have a better insight into the classifier' performance, three additional classification metrics other than accuracy (ACC) are calculated as well. These metrics are precision (PRE), recall (REC), $F_1$ score ($F_1$). These metrics are defined for a binary classification problem as follows:

$$PRE = TP/(TP + FP) \qquad (1)$$

$$REC = TP/(TP + FN) \qquad (2)$$

$$F_1 = 2 \times (PRE \times REC)/(PRE + REC) \qquad (3)$$

where TP is True Positive, which is the number of events that are correctly predicted as the true class; FP is False Positive, which is the number of events that are incorrectly predicted as true class; FN is False Negative, which is the number of events that are incorrectly predicted as not be true class; and TN is True Negative, which is the number of events that are correctly predicted as not be the true class. The metrics in the multiclass problem is still the same as the ones used in the binary classification; however, the metrics are calculated for each class by treating it as a binary classification problem after combining all non-true classes into the second class. Then, the binary metrics are averaged over all the classes to get either a macro average (treat each class equally) or weighted average (weighted by class frequency) metric [25].

Table. I shows the event classification with respect to the sampling rates. Two sampling rates with 30 fps and 60 fps are considered with input feature of ROCOF and the accuracy and weighted average of the precision, recall, and $F_1$ score are compared with each other.

As shown in Table I, the classification metrics improve as the sampling rate increases. The classification metrics go up from about 50% to 60% for 30 fps to close to 90% for 60 fps sampling rate, accounted for more than 30% improvement. This clearly shows that a higher sampling rate results in better classification performance.

TABLE I. SAMPLING RATE ANALYSIS

| Sampling Rate | ACC (%) | PRE (%) | REC (%) | $F_1$ (%) |
|---|---|---|---|---|
| **30 (fps)** | 60 | 49 | 60 | 51 |
| **60 (fps)** | 88 | 89 | 88 | 88 |

Fig. 6 (left) and (right) show the confusion matrix with 30 fps and 60 fps sampling rates, respectively. The confusion matrix shows the performance of the proposed event classification method for distinguishing the true events versus the misclassified ones. The rows show the true class and the columns correspond to the predicted class. The diagonal entries correspond to events that are correctly predicted, and the off-diagonal entries correspond to events that are incorrectly predicted. The number of events of the total number of events in each case is shown in each cell.

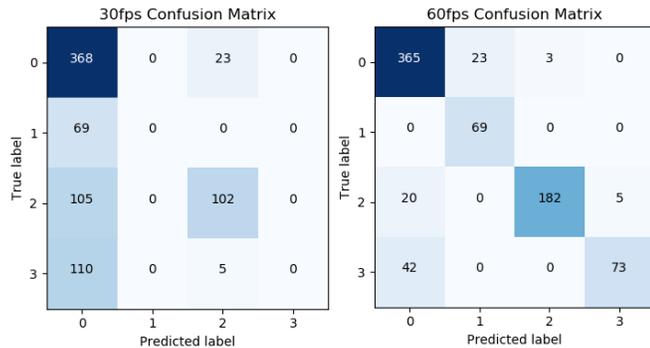

Fig. 6. Confusion matrix (left) 30 fps (right) 60 fps

It can be seen that using the 30 fps PMU data, the classifier has only good recall in classifying the Line Outage event (class 0) and about 50% success in distinguishing the Frequency event (class 2). However, distinctly using 60 fps PMU data, the model has perfect recall for Transformer Outage (class 1). It successfully classifies all the 69 instances of this class. While for the model trained with 30 fps data the recall was zero. In addition, the precision and recall for Oscillation event (class 3) are jumped from zero to about 90%, and 60%, respectively.

V. CONCLUSION

This paper presents a wide-area event classification in transmission power grids. The dataset for this paper is the real-world dataset of anonymized PMUs data from one of the main U.S. transmission interconnections. The event classification is formulated using deep neural networks (DNN) to learn four major events in the dataset including line outage, transformer outage, frequency event, and oscillation event. For tuning the hyperparameters of the DNN and to guarantee the best model performance, Bayesian optimization based hyperparameter tuning is implemented. The effectiveness of the proposed framework is validated through several case studies and for different PMU data, such as gradient of voltage, current or frequency. For future work, we intend to apply the models on a much larger dataset over the course of two years. Based on the results concluded from the size of the training dataset, it is expected that the DNN model performance would improve with the increase of the training dataset size. In addition, the possibility of using different optimization algorithms for tuning the neural networks will be considered.